\documentclass[preprint2]{aastex}

\usepackage{graphicx,subfigure}
\usepackage{pslatex}

\newcommand{\myemail}{khalack.viktor@umoncton.ca}

\shorttitle{Project VeSElkA: Analysis of Balmer line profiles}
\shortauthors{Khalack \& LeBlanc}

\begin{document}

\title{Project VeSElkA: Analysis of Balmer line profiles in slowly rotating chemically peculiar stars\protect\thanks{Based on observations obtained at the Canada-France-Hawaii Telescope (CFHT) which is operated by the National Research Council of Canada, the Institut National des Sciences de l'Univers of the Centre National de la Recherche Scientifique of France, and the University of Hawaii.}}

\author{V. Khalack and F. LeBlanc}
\affil{D\'epartement de Physique et d'Astronomie, Universit\'e de Moncton, Moncton, N.-B., Canada E1A 3E9 }
\email{\myemail}

%\label{firstpage}

\begin{abstract}
We present results for the estimation of gravity, effective temperature and radial velocity of poorly studied chemically peculiar stars recently observed with the spectropolarimeter ESPaDOnS at CFHT in the frame of the VeSElkA (Vertical Stratification of Element Abundances) project.
For four of the stars from our sample (\objectname{HD~23878}, \objectname{HD~83373}, \objectname{HD~95608} and \objectname{HD~164584}), the values of their effective temperature and surface gravity are determined for the very first time.
Grids of stellar atmosphere models with the corresponding fluxes have been calculated using version 15 of the PHOENIX code for effective temperatures in the range of 5000K to 15000K, for the logarithm of surface gravities in the range of 3.0 to 4.5 and for the metallicities from -1.0 to +1.5. We have used these fluxes to fit the Balmer line profiles employing the code FITSB2 that produces estimates of the effective temperature, gravity and radial velocity for each star.
When possible, our results are compared to those previously published.
The physical characteristics of 16 program stars are discussed with the future aim to study the abundance anomalies of chemical species and the possible vertical abundance stratification in their stellar atmosphere.

\end{abstract}


\section{Introduction}

Despite of the fact that some chemically peculiar (CP) stars have "stable" atmospheres, they sometimes also show variability of their spectra with the period of stellar rotation due to the horizontal inhomogeneous distribution of elements abundance in their stellar atmosphere \citep{Khokhlova75}. A significant fraction of CP stars shows signatures of strong magnetic fields \citep{Bychkov+03} and their structure correlates with the patches of overabundance (or underabundance) of some elements \citep{Kochukhov+02, Shavrina+10}. It appears that the magnetic field can intensify accumulation or depletion of chemical elements at certain optical depths \citep{Ryab+08, Alecian+Stift10}. Some stars also show vertical stratification of abundance of several chemical species \citep{Savanov+Kochukhov98, Ryab+04, Khalack+13}, which can be explained in terms of the mechanism of atomic diffusion \citep{Michaud70}.
%It appears that the light and the iron-peak elements are concentrated in the lower atmospheric layers (Ryabchikova et al. 2004), while the rare-earth elements (for example, Pr and Nd) are pushed into the upper atmosphere (Mashonkina et al. 2005).

Accumulation or depletion of chemical elements at certain optical depths brought about by atomic diffusion can modify the structure of stellar atmospheres (Hui-Bon-Hoa et al. 2000; LeBlanc et al. 2009, 2010) and it is, therefore, important to gauge the intensity of such stratification. The slowly rotating CP stars with $V\sin{i} <$ 40 km s$^{-1}$ are good candidates to study the abundance stratification of elements with optical depth. Their small rotational velocities result in comparatively narrow and unblended line profiles which are well suitable for abundance analysis. Based on their slow rotation, we may assume the hypotheses of a hydrodynamically stable atmosphere in these stars which is necessary for the diffusion process to take place.

\begin{table}[t]
%\parbox[t]{\textwidth}{
\centering
\caption[]{List of the observed slowly rotating CP stars.}
\begin{tabular}{l|crr}
\hline
\hline
     &           &              &      \\
Star & $m_{\rm V}$ & $\Delta t$ & S/N  \\
     &           &   (s)        &      \\
%\multicolumn{2}{c}{Energy, cm$^{-1}$ (NIST)}\\ \cline{2-6}
\hline
\objectname{HD~15385} &	6.2	& 1600 & 1100 \\ % & K\"{u}nzli \& North (1998)\\
\objectname{HD~22920} &	5.5	&  920 & 1150 \\ % & Catanzaro et al. (1999) \\
\objectname{HD~23878} &	5.2	&  660 &  950 \\ % &   \\
%\objectname{HD~24712} &	6.0	& 1280 & 1200 \\
\objectname{HD~53929} &	6.1 & 1160 & 1000  \\ % & Smith \& Dworetsky (1993)\\
\objectname{HD~68351} &  5.6	& 920  & 1100  \\ % & \\
\objectname{HD~71030} &  6.1	& 1140 & 1100  \\ % & Prugniel et al. (2011) \\
\objectname{HD~83373} &  6.4	& 1524 & 1000   \\ % & \\
\objectname{HD~90277} &  4.7	& 520  & 1000  \\ % & Berthet (1990)\\
\objectname{HD~95608} &  4.4	& 416  & 1300    \\ % & \\
\objectname{HD~97633} &  3.3	& 152  & 1300  \\ % & Koleva \& Vazdekis (2012) \\
\objectname{HD~110380}&	3.6	& 200  & 1300  \\ % & \\
\objectname{HD~116235}&	5.9	& 1040 &  870  \\ % & \\
\objectname{HD~164584}&	5.4	& 880  & 1300  \\ % & \\ %\hline
%\objectname{HD~170973}&	6.4	& 2000 & 1060 &10000$\pm$200 & 3.01$\pm$0.1&-10.5$\pm$1.0& 0.88 & 12046$\pm$925$^b$ & 3.56$\pm$1.74$^b$  & 18$^b$    \\ % & Koleva \& Vazdekis (2012) \\
\objectname{HD~186568}& 6.0 & 1300 & 1000 \\
\objectname{HD~209459}&	5.8	& 1160 & 1000   \\ % & Prugniel et al. (2011) \\
\objectname{HD~223640}&	5.2	&  680 & 1200   \\ % & Prugniel et al. (2011) \\
\hline
\end{tabular}
\label{tab1}
\end{table}

The diffusion velocities of different chemical species depend on the relative values of gravity and radiative acceleration resulting from the momentum
transfer from the radiation field to these chemical species (e.g. Gonzalez et al. 1995). The momentum transfer depends on the opacity of the species under consideration and on the local monochromatic radiation field, which, in turn, via the monochromatic opacities, depends on the local abundances of the different species. In a hydrodynamically stable atmosphere, the diffusion process may result in vertical stratification of chemical abundances.
%the elements abundance that satisfies the conditions of dynamical equilibrium between the gravitational and radiative forth, when the diffusion velocity of each element is nil at each layer of the atmosphere.
Detection of vertical abundance stratification of chemical species in atmospheres of slowly rotating CP stars is an indicator of the effectiveness of the diffusion mechanism  responsible for the observed peculiarities of chemical abundances. Comparing the observed characteristics of vertical stratification of chemical abundances with the results of theoretical modeling will help to verify and improve the self-consistent models of stellar atmospheres. Such models were calculated with a modified version of the PHOENIX code \citep{LeBlanc+09}. These models were mostly applied to blue horizontal-branch stars \citep{LeBlanc+10}. \citet{Stift+Alecian12} also have calculated chemically stratified atmospheric models  using the CARATSTRAT code applied to ApBp stars.

Vertical stratification of the chemical abundances in stars can be estimated through the analysis of multiple lines that belong to the same ion of the studied element \citep{Khalack+Wade06, Khalack+07} using the ZEEMAN2 code \citep{Landstreet88}. This procedure was successfully implemented to detect vertical abundance stratification in the atmospheres of several blue horizontal-branch stars (Khalack et al. 2007, 2008, 2010).

%The data on abundance variation with atmospheric depth for different chemical species and the magnetic field measurements obtained for a particular rotational phase will be combined with the already published data to verify how they are changing with the stellar axial rotation. A change of the vertical abundance stratification can be detected from the comparative analysis of several sets of Stokes IV spectra acquired for different rotational phases. Those changes can be caused by the configuration of magnetic field that modifies atomic diffusion and, therefore, the abundance stratification of certain elements.

Therefore, we have initiated a new project entitled "Vertical Stratification of Elements Abundances" (VeSElkA - meaning rainbow in Ukrainian) aimed to search for, and study the signatures of abundance stratification of chemical species with optical depth in the atmospheres of slowly rotating CP stars. A list of suitable candidates has been compiled based on the catalog of Ap, HgMn and Am stars of \citet{Renson+Manfroid09}. From the very beginning, we have concentrated our attention on the relatively bright, slowly rotating and poorly studied CP stars that can be easily observed with ESPaDOnS (CFHT) in spectropolarimetric mode. Up to now, we have obtained high signal-to-noise ratio and high resolution spectra for a sample of selected CP stars (see Sec.~\ref{obs}) and found indication of vertical stratification of iron abundance in \objectname{HD~95608} and \objectname{HD~116235} \citep{Khalack+13}. % and \objectname{HD~22920} \citep{Khalack+Poitras14}.
Not all stars in our sample are expected to contain vertical stratification of elements in their atmosphere.
For instance, AmFm stars are expected to have chemically homogeneous atmospheres due to convective mixing \citep{Richer+00, Richard+01}. %(Richer et al. 2000; Richard et al. 2001).
Our analysis will however give estimates for the average abundances in the atmospheres of these stars. Such results could help confirm or disprove their classification as CP stars.
%Recently, we have obtained the ESPaDOnS spectra of several slowly rotating CP stars with the aim to study a vertical stratification of chemical species in stellar atmosphere.

This paper aims to present results for the determination of the fundamental parameters ($T_{\rm eff}$ and $\log{g}$) as well as radial velocity of 16 selected stars. This is the first estimation of $T_{\rm eff}$ and $\log{g}$ for four of them. %The abundance analysis for these stars is underway and will be presented in future publications.
The observations and the reduction procedure are described in Section~\ref{obs}. The models used and the fitting procedure are discussed in Sections~\ref{grid} and~\ref{fit} respectively. The main results are presented in Section~\ref{stars} together with the description of the properties of each studied star. A discussion follows in Section~\ref{summary}.

%Star	mV	Δt, s.	S/N	Teff, K	log (g)	Vsin(i) km/s

%For several CP stars, some spectral data are available at the astronomical archives, but I also plan to obtain additional data with the spectropolarimeter ESPaDoNS at CFHT (two applications for the Semester 2013A have already been submitted to CFHT). For some short-periodic (P < 13d) and slowly rotating magnetic CP stars, I plan to accumulate spectropolarimetric (Stokes IVQU) data that will be obtained for different rotational phases. From the combined analysis of Balmer lines and some HeI lines, my students and I will obtain the effective temperature and gravity for the studied stars employing the code FITPROF 2.4 (Napiwotzki et al. 1999). The maps of horizontal abundance distribution are going to be constructed using the Magnetic Doppler Imaging (MDI) technique (Piskunov \& Kochukhov 2002). Application of the MDI technique provides information on the structure of magnetic field, which modifies Stokes IVQU profiles due to the Zeeman and in some cases due to the partial Paschen-Back (hereafter PPB) effect. This study aims to increase the precision of abundance mapping and reconstruction of magnetic field configuration. Contrarily to the previous studies, where only the Zeeman effect was used, we will also employ the PPB effect (that I recently implemented in the ZEEMAN2 code, see Khalack \& Landstreet (2012)) to simulate the magnetically sensitive line profiles in the spectra of Ap stars with strong magnetic fields. The abundance maps will be analyzed for correlation with the recovered magnetic field structure.

\begin{figure*}[t]
\includegraphics[width=3.25in,angle=0]{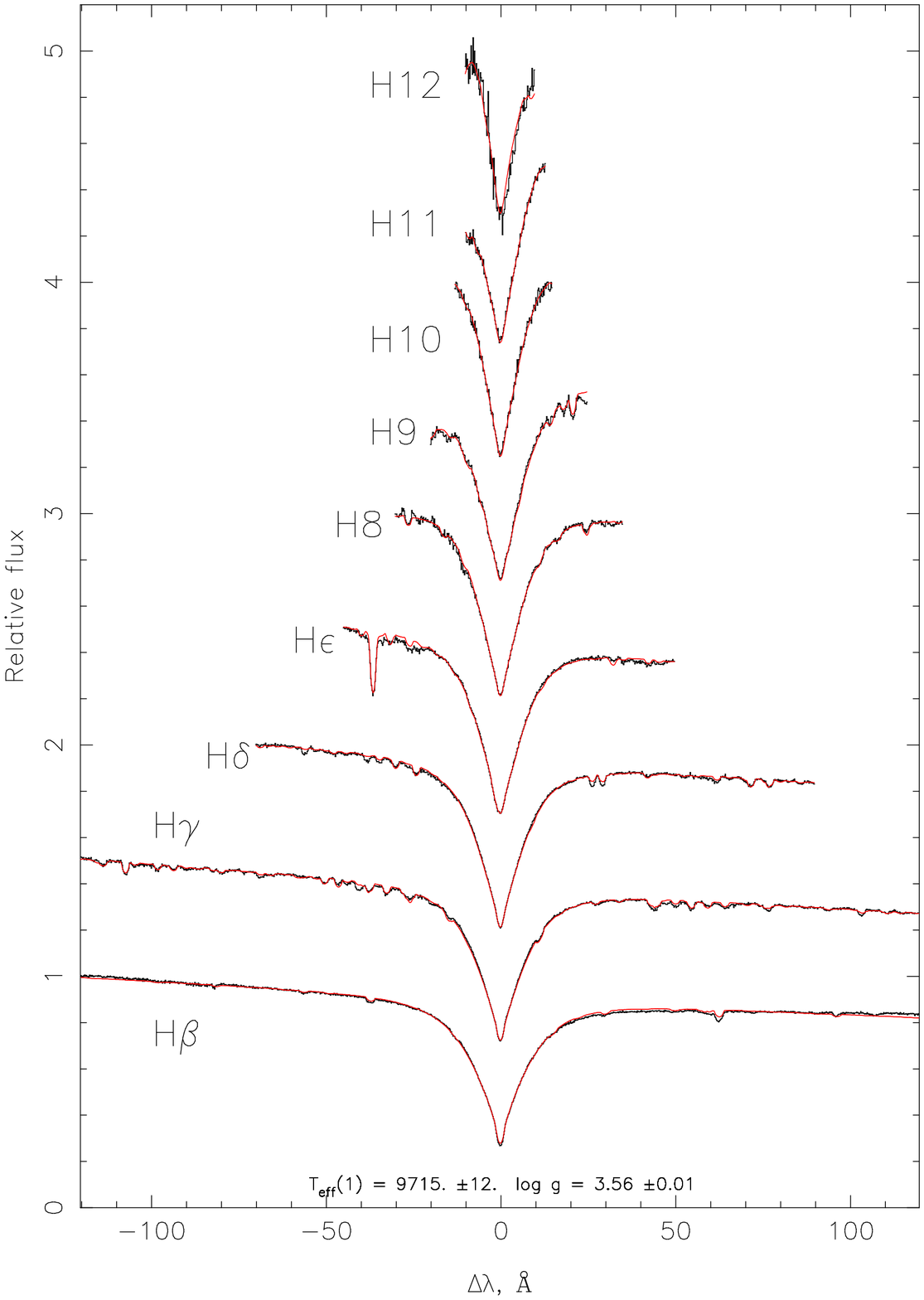}
\includegraphics[width=3.25in,angle=0]{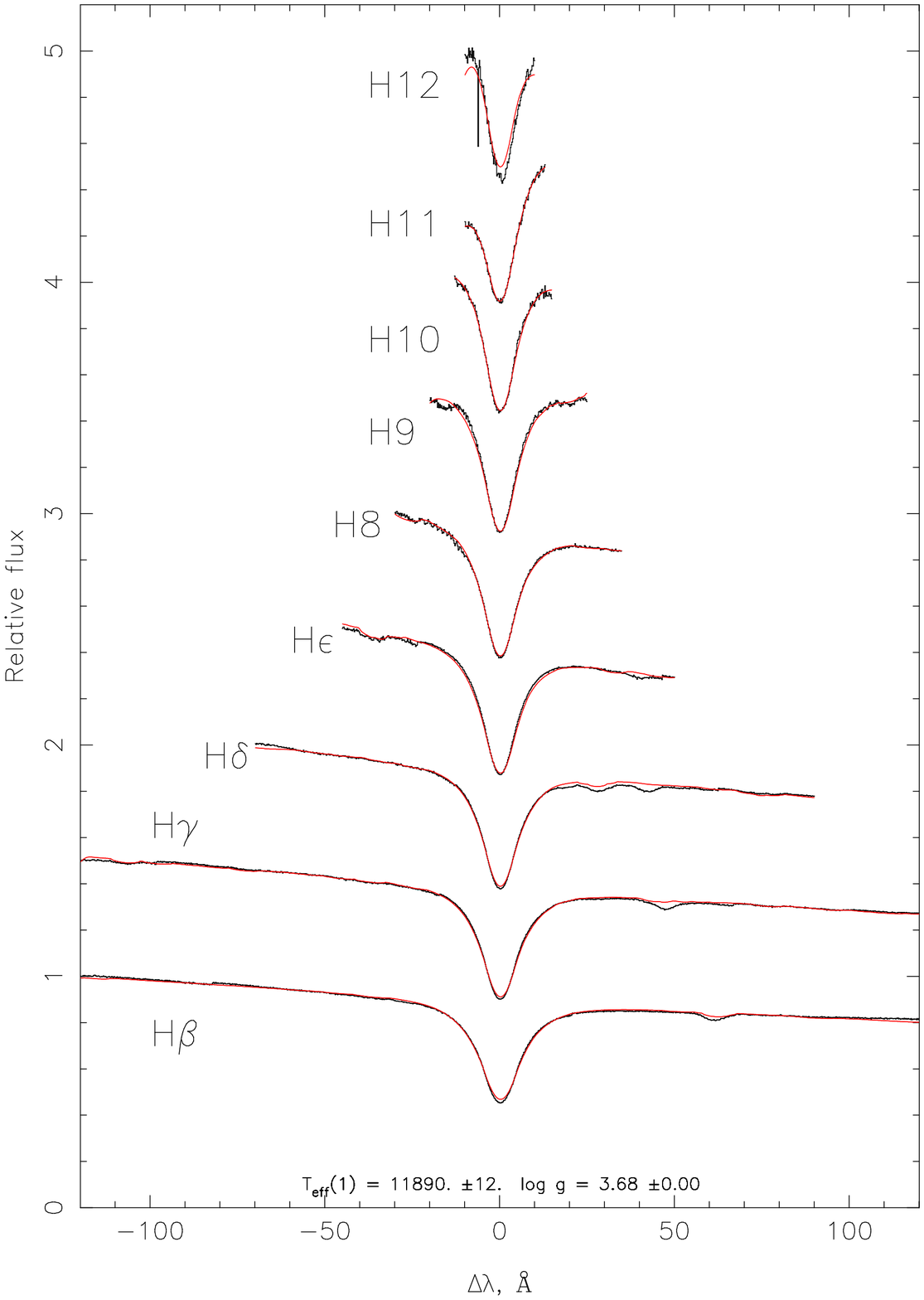}
\caption{ The observed spectra (thick line) of 108 Vir (left) and $\alpha$ Leo (right) are well fitted by synthetic spectra (thin dotted line) that corresponds to $T_{\rm eff}$ = 9715 K, $\log{g}$ = 3.56, [M/H]= -0.5 ($\chi^2/\nu$ = 2.64) and $T_{\rm eff}$ = 11890 K, $\log{g}$ = 3.68, [M/H]= 0.0 ($\chi^2/\nu$ = 12.26) respectively. }
\label{fig1}
\end{figure*}

\section{Observations}
\label{obs}
High resolution (R=65000) Stokes IV spectra of several CP stars with $V\sin{i} <$ 40 km s$^{-1}$ were obtained recently with ESPaDOnS
(Echelle SpectroPolarimetric Device for Observations of Stars) employing the deep-depletion e2v device Olapa.
ESPaDOnS allows the acquisition of an essentially continuous spectrum throughout the spectral range from 3700\AA\, to 10500\AA\, in a single exposure \citep{Donati+99}.
The optical characteristics of the spectrograph as well as the instrument performances are described by \citet{Donati+06}\footnote{
For more details about this instrument, the reader is invited to visit {\rm www.cfht.hawaii.edu/Instruments/Spectroscopy/Espadons/}}.
The obtained spectra were reduced using the dedicated software package Libre-ESpRIT \citep{Donati+97}
which yields both the Stokes I spectrum and the Stokes V circular polarisation spectrum.
To infer the effective temperature and gravity of the observed stars, we have used their non-normalized spectra.

Table~\ref{tab1} presents a list of the observed slowly rotating chemically peculiar stars. The first and the second columns provide respectively the name of the star and its apparent visual magnitude, while the third and the forth columns contain the accumulation time and the maximal signal-to-noise ratio for the acquired Stokes I spectra.

%\section{Determination of effective temperature and gravity}
%\label{tlogg}

%\subsection{Stellar atmosphere models}
%\label{models}

\section{Grid of models}
\label{grid}

A new library of high resolution synthetic spectra has been created in order to determine the effective temperature and gravity of stars observed in the frame of the
project VeSElkA. A grid of stellar atmosphere models and corresponding fluxes have been calculated using version 15 of the PHOENIX code \citep{Hauschildt+97} for  5000 K $\leqslant T_{\rm eff} \leqslant$ 15000 K and 3.0 $\leqslant \log{g} \leqslant$ 4.5. For the effective temperatures from 5000 K to 9000 K, we have used a 250 K step, while for the higher temperatures up to 15000 K a 500 K step was used. The gravities have been calculated with a step 0.5. The theoretical fluxes have a resolution of 0.05\AA\, in the visible range from 3700\AA\, to 7700\AA. We have calculated grids of models for solar metallicity \citep{Grevesse+10} as well as for the metallicities [M/H]= -1.0, -0.5, +0.5, +1.0, +1.5. For all the grids, the microturbulent velocity is assumed to be zero. The synthetic spectra have been corrected for the air wavelength using the equation given by \citet{Morton91}.

Exploiting the possibilities of the PHOENIX code \citep{Hauschildt+97}, we have calculated spherically symmetric models of stellar atmospheres for which the gravity depends on the stellar radius. In this case, all the structure of the atmosphere can be calculated knowing
%important physical parameters of a stars can be derived from
the effective temperature $T_{\rm eff}$, the surface gravity $\log{g}$ and the stellar mass $M_{\rm *}$. For the given values of effective temperature and surface gravity, we have derived the respective stellar mass trough the interpolation of data for physical parameters of normal main-sequence stars \citep{Popper80}.

\begin{figure*}[t]
\includegraphics[width=2.25in,angle=-90]{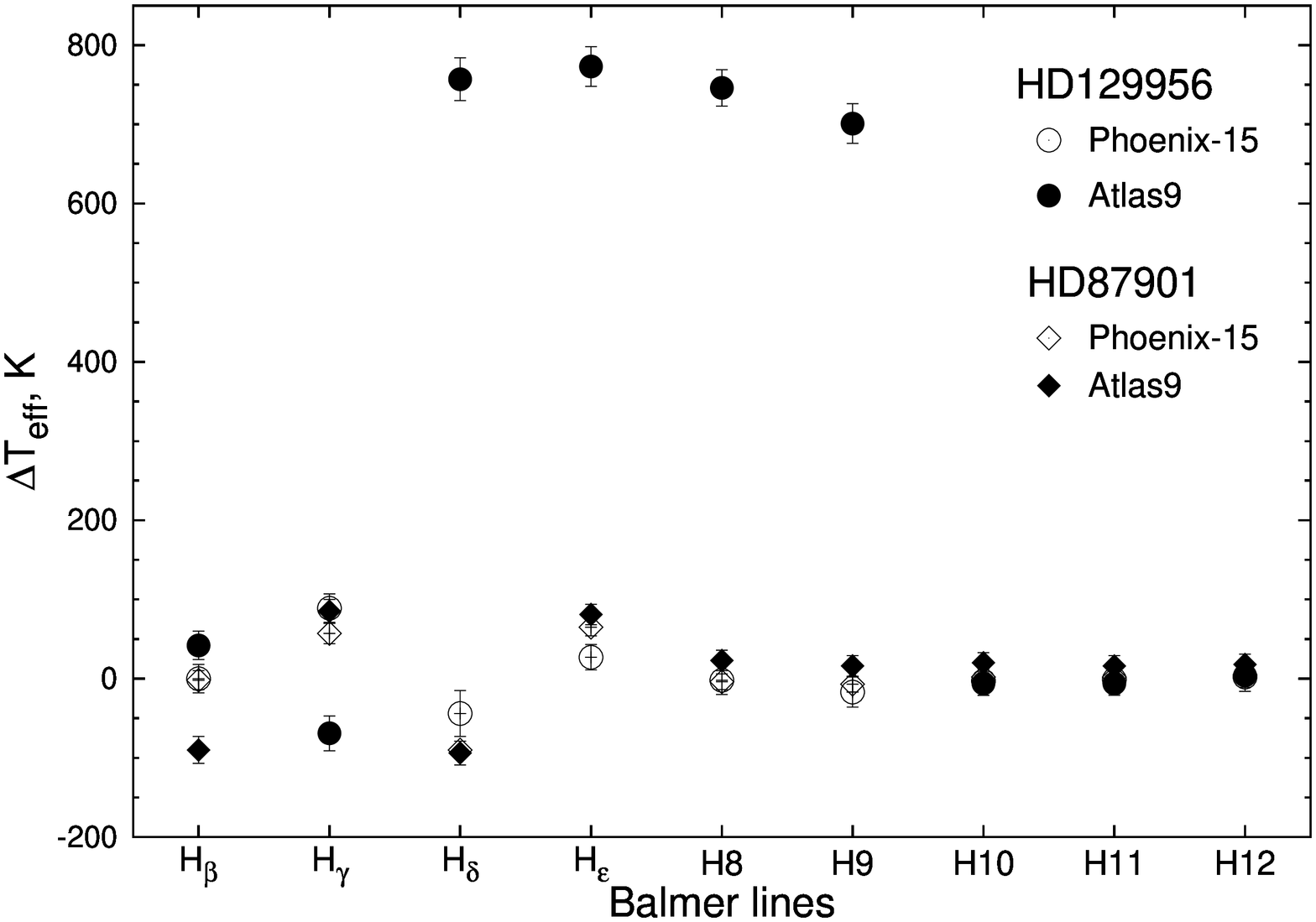}
\includegraphics[width=2.25in,angle=-90]{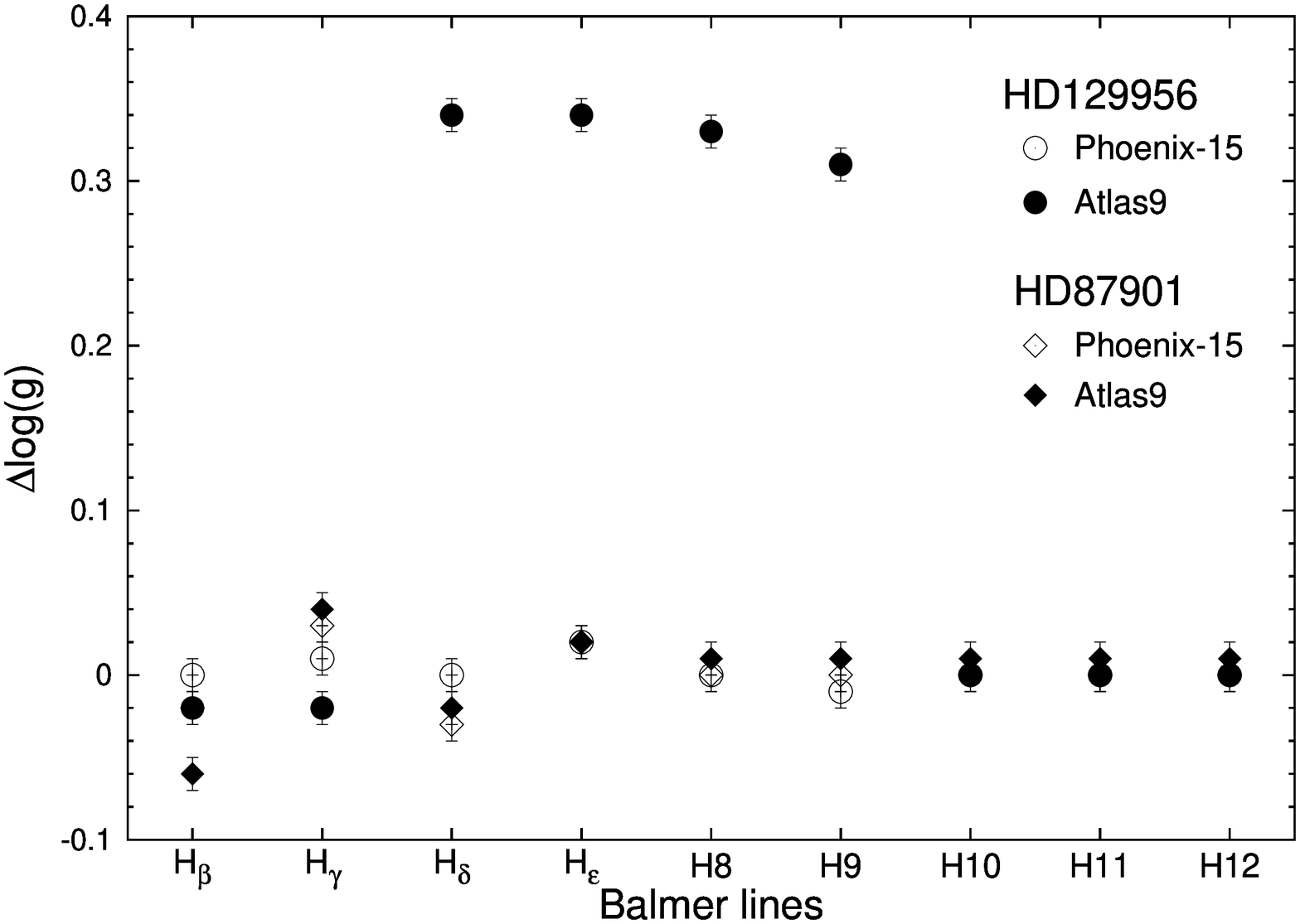}
\caption{ Variation of the best fit results obtained, when one of the Balmer lines (indicated at the X-axis) is omitted from the analysis, with respect to the values of $T_{\rm eff}$ (left) and $\log{g}$ (right) obtained from the analysis of all 9 Balmer line profiles (see Fig.~\ref{fig1}) for HD~129956 (108~Vir) using the Phoenix-15 grids (open circles) and the Atlas9 grids (filled circles), and for HD~87901 ($\alpha$ Leo) using the Phoenix-15 grids (open diamonds) and the Atlas9 grids (filled diamonds). The error bars represent the internal errors of the fitting procedure with the FITSB2 code and are much smaller than the true errors (see Section~\ref{grid}).}
\label{fig2a}
\end{figure*}

To verify the accuracy of the grids of stellar atmosphere models, we have used the spectra of %Vega \citep{Petit+10}
$\alpha$ Leo (HD~87901) and 108 Vir (HD~129956)\footnote{Spectra of 108 Vir and of $\alpha$ Leo where downloaded from the CADC database via http://www4.cadc-ccda.hia-iha.nrc-cnrc.gc.ca/en/cfht/.} obtained with ESPaDOnS in the spectropolarimetric mode. The profiles of nine Balmer lines in the non-normalised spectra of $\alpha$ Leo and 108 Vir have been fitted with the help of the FITSB2 code \citep{Napiwotzki+04} using the grid of models calculated for the metallicity [M/H]=-0.5 and 0.0 respectively (see Fig.~\ref{fig1}). In order to model a continuum in the vicinity of the analysed Balmer lines, the code FITSB2 uses a linear function fitted over the line profile region and the continuum as a part of the fitting procedure. This approach provides a better fit quality than the standard procedure in which one tries to fit the continuum regions (relatively free of lines) to the left and to the right of the analysed Balmer line profile separately \citep{Napiwotzki14}.

%{\bf In the case of Vega}, the best fit is obtained for $T_{\rm eff}$ = 9690 K and $\log{g}$ = 4.18, resulting in $\chi^2/\nu$ = 2.01. Exactly the same approach has been applied to fit the Balmer lines in the same spectrum using the grid of models calculated with version 16 of the PHOENIX code for the metallicity [M/H]=-0.5 \citep{Husser+13}. The best fit has resulted in $T_{\rm eff}$ = 9630 K and $\log{g}$ = 4.14 ($\chi^2/\nu$ = 2.0). These values of effective temperature and surface gravity are close to one another and to the values $T_{\rm eff}$ = 9560 K and $\log{g}$ = 4.05 obtained for Vega by \citet{H+L+93} using the results of $ubvyH_{\beta}$ photometry. {\bf Comparison with ATLAS9 models?? Vega is very rapidly rotating star with a substantially inhomogeneous local effective temperature (e.g., Hill et al. 2010, ApJ 712, 250)}

In the case of $\alpha$ Leo, assuming its $V\sin{i}$ = 300 km s$^{-1}$ \citep{Abt+02}, the best fit is obtained for solar metallicity, $T_{\rm eff}$ = 11890 K and $\log{g}$ = 3.68, resulting in $\chi^2/\nu$ = 12.26 (see right panel of the Fig.~\ref{fig1}). A similar simulation that employs the grid of models calculated with PHOENIX-16 code \citep{Husser+13} gives the best fit for the solar metallicity, $T_{\rm eff}$ = 11976 K and $\log{g}$ = 3.75 ($\chi^2/\nu$ = 11.57). These results are close to each other and to the values $T_{\rm eff}$ = 11962 $\pm$ 115 K and $\log{g}$ = 3.56 %$\pm$ 0.1
obtained by \citet{Gray+03} from fitting low resolution spectra of $\alpha$ Leo and its photometric data. Meanwhile, using the Atlas9 grids\footnote{The Atlas9 grids are available at the http://zuserver2.star.ucl.ac.uk/~idh/NewGrids/Atlas9.C04/.} \citep{C+K04} to fit the Balmer line profiles in the same spectrum of $\alpha$ Leo, the best fit is obtained for the metallicity [M/H] = -0.3, $T_{\rm eff}$ = 12660 K and $\log{g}$ = 3.80 ($\chi^2/\nu$ = 13.76).  It appears that for $\alpha$ Leo the use of the Atlas9 grids leads to the best fit with comparatively higher value of $\chi^2/\nu$ and to higher values of effective temperature and surface gravity.

In the case of 108 Vir, assuming its $V\sin{i}$ = 83 km s$^{-1}$ \citep{Ammler+Reiners12}, the best fit is obtained for the metallicity [M/H]=-0.5, $T_{\rm eff}$ = 9715 K and $\log{g}$ = 3.56, resulting in $\chi^2/\nu$ = 2.64 (see left panel of Fig.~\ref{fig1}). The grids calculated with PHOENIX-16 code and the Atlas9 grids result in the best fits with [M/H] = -0.2, $T_{\rm eff}$ = 10280 K, $\log{g}$ = 3.86 ($\chi^2/\nu$ = 2.57) and [M/H] = -0.3, $T_{\rm eff}$ = 9760 K, $\log{g}$ = 3.64 ($\chi^2/\nu$ = 2.88) respectively. Our estimate of the effective temperature is close the one obtained with the Atlas9 grids and to $T_{\rm eff}$ = 9840 K found by \citet{Ammler+Reiners12}, but is significantly smaller than the effective temperature obtained using the PHOENIX-16 grids \citep{Husser+13}. Our estimate of the surface gravity is similar to the one obtained for the Atlas9 grids but smaller than the one obtained with the PHOENIX-16 grids.

%To verify the accuracy of the grids of our stellar atmosphere models, we have also used another spectroscopic standard, HR~718, for which \citep{Ammler+Reiners12} found $T_{\rm eff}$ = 10396 K and $V\sin{i}$ = 57.6 km s$^{-1}$. ????

By comparing our results for $\alpha$ Leo and 108 Vir with the relatively well-established values for $T_{\rm eff}$ and $\log{g}$, we may roughly estimate the uncertainties of the results presented here as  $\pm$200 K for effective temperature and $\pm$0.2 for surface gravity.
%{\bf Part of the errors may be related to the larger step for the effective temperature used in our grids in comparison with the PHOENIX-16 grids \citep{Husser+13} and Atlas9 grids \citep{C+K04}. }
%We can deduce from the comparison of these results that application of our grids of models calculated with PHOENIX (ver. 15) provide a precision of approximately $\pm$200 K for effective temperature and $\pm$0.2 for surface gravity.

%The Table~\ref{tab1} contains a list of the observed chemically peculiar stars. The first and the second columns present a name of the star and its apparent visual magnitude respectively, while the third and the forth columns provide give the accumulation time and the maximal signal-to-noise ratio for the acquired spectra.

%two other spectroscopic standards, HR~3982 and 108 Vir, that have Teff close to 10000K. In the case of 108 Vir the agreement for Teff=9715K (Phoenix-15:) and Teff=9840K (Ammler-von Eiff \& Reiners 2012, A\&A, 542, 116) is good (we have not found published data for log(g) in 108 Vir). Meanwhile, in the case of HR~3982 the difference between Teff=10068K (Phoenix-15:) and Teff=10396K (Ammler-von Eiff \& Reiners 2012) is more than 300K.

\begin{table*}
%\parbox[t]{\textwidth}{
\centering
\caption[]{Determined values of the effective temperature and the surface gravity for the programm CP stars.}
\begin{tabular}{l|rccrc|ccc }
\hline
\hline
     &  \multicolumn{5}{c|}{Balmer lines}& \multicolumn{3}{c}{Previous publications}  \\
Star &  $T_{\rm eff}$ & $\log{g}$ & $V_{\rm r}$ & [M/H] & $\chi^2/\nu$ & $T_{\rm eff}$ & $\log{g}$& $V\sin{i}$     \\
     &     (K)         &          & (km s$^{-1}$) &     &              &  (K)          &          & (km s$^{-1}$)  \\
%\multicolumn{2}{c}{Energy, cm$^{-1}$ (NIST)}\\ \cline{2-6}
\hline
\objectname{HD~15385} & 8230$\pm$200 & 4.00$\pm$0.2& 22.0$\pm$1.0& 0.0& 0.65 &  8154$^a$ & 4.12$^a$ & 21$^a$, 29$^b$ \\ % & K\"{u}nzli \& North (1998)\\
\objectname{HD~22920} &13640$\pm$200 & 3.74$\pm$0.2& 20.0$\pm$2.0&-0.5& 5.67 & 13700$^c$ & 3.72$^c$ & 30$^d$, 39$^b$  \\ % & Catanzaro et al. (1999) \\
\objectname{HD~23878} &	8740$\pm$200 & 3.86$\pm$0.2& 29.5$\pm$1.0& 0.0& 0.62 &   &   & 24$^b$  \\ % &   \\
%\objectname{HD~24712}  &	7370$\pm$100 & 3.76$\pm$0.1& 22.0$\pm$1.0& 0.75 & 7250$\pm$150$^h$ & 4.20$\pm$0.10$^h$ & 18$^b$  \\ % & Ryabchikova et al. (1997) \\
\objectname{HD~53929} &13950$\pm$200 & 3.90$\pm$0.2& 15.0$\pm$1.0&-1.0& 3.20 & 14050$\pm$250$^e$& 3.60$\pm$0.25$^e$	& 25$^b$, 30$^e$  \\ % & Smith \& Dworetsky (1993)\\
\objectname{HD~68351} &10080$\pm$200 & 3.22$\pm$0.2&18.1$\pm$1.0& 0.0&  1.17 & 10290$\pm$340$^f$&    & 33$^b$  \\ % & \\
\objectname{HD~71030} &	6780$\pm$200 & 4.04$\pm$0.2& 38.1$\pm$1.0$^h$& 0.0& 0.28 & 6541$\pm$47$^g$ & 4.03$\pm$0.05$^g$ & 9$\pm$2$^h$  \\ % & Prugniel et al. (2011) \\
\objectname{HD~83373} &	9800$\pm$200 & 3.81$\pm$0.2& 26.5$\pm$1.0& 0.0& 0.92 &   &  & 28$^b$  \\ % & \\
\objectname{HD~90277} &	7250$\pm$200 & 3.62$\pm$0.2& 14.5$\pm$1.0& 0.0& 1.29 & 7440$^i$ & 3.46$^i$ & 34$^b$  \\ % & Berthet (1990)\\
\objectname{HD~95608} &	9200$\pm$200 & 4.25$\pm$0.2&-10.4$\pm$1.0$^h$&+0.5& 0.59 &   &   & 21$^b$,17$\pm$2$^h$     \\ % & \\
\objectname{HD~97633} &	8750$\pm$200 & 3.45$\pm$0.2&  8.2$\pm$1.0& 0.0& 0.61& 8790$\pm$351$^j$ & 3.59$\pm$0.89$^j$ & 23$^b$  \\ % & Koleva \& Vazdekis (2012) \\
\objectname{HD~110380}&	6980$\pm$200 & 4.19$\pm$0.2&-17.6$\pm$1.0& 0.0& 0.31& 6720$^k$  & 4.20$^k$  & 23$^b$  \\ % & \\
\objectname{HD~116235}&	8900$\pm$200 & 4.33$\pm$0.2&-10.3$\pm$1.0$^h$&+0.5& 0.49 & 8570$^l$ & 4.23$^l$ & 20$\pm$2$^h$ \\ % & \\
\objectname{HD~164584}&	6800$\pm$200 & 3.54$\pm$0.2&-11.2$\pm$1.0& 0.0& 1.16 &   &   & \\ % & \\ %\hline
%\objectname{HD~170973} &10000$\pm$200 & 3.01$\pm$0.1&-10.5$\pm$1.0& 0.88 & 12046$\pm$925$^b$ & 3.56$\pm$1.74$^b$  & 18$^b$    \\ % & Koleva \& Vazdekis (2012) \\
\objectname{HD~186568}&11070$\pm$200 & 3.44$\pm$0.2& -9.5$\pm$1.0& -0.5& 1.81 & 11596$\pm$120$^m$ & 3.39$\pm$0.15$^m$ & 18$^m$ \\
\objectname{HD~209459}&10310$\pm$200 & 3.62$\pm$0.2& -0.3$\pm$1.0& 0.0& 0.97 & 10455$\pm$400$^m$ & 3.52$\pm$0.15$^m$ & 14$^b$  \\
\objectname{HD~223640}&12500$\pm$200 & 4.08$\pm$0.2& 17.0$\pm$2.0&+1.0& 1.65 & 12429$\pm$435$^g$ & 3.93$\pm$0.23$^g$ & 28$^b$  \\ % & Prugniel et al. (2011) \\
\hline
\end{tabular}
\label{tab1b}
%}
\scriptsize{
 \\ {\it Notes:} $^a$\citet{Kunzli+North98}, $^b$\citet{Royer+02}, $^c$\citet{Catanzaro+99}, $^d$\citet{L+M96}, $^e$\citet{Smith+Dworetsky93}, $^f$\citet{Auriere+07}, $^g$\citet{Prugniel+11}, $^h$\citet{Khalack+13}, $^i$\citet{Berthet90}, $^j$\citet{Koleva+Vazdekis12}, $^k$\citet{B+T86}, $^l$\citet{Erspamer+North03}, $^m$\citet{Hubrig+Castelli01} }
 %$^h$Ryabchikova et al. \citet{Ryab+97},
 %$^j$Wielen et al. \citet{Wielen+00}, $^k$Wielen \citet{Wielen99},
 %$^l$Evans \citet{Evans67}, $^m$Nordstr\"{o}m et al. \citet{Nordstreom+04}, $^n$Bailey \& Landstreet \citet{Bailey+Landstreet13}, $^o$Gontcharov \citet{Gontcharov2006}, $^p$Wilson \citet{Wilson53}}, $^r$Leone \& Manfr\`{e} \citet{L+M96},
\end{table*}

\section{Fitting procedure}
\label{fit}
\subsection{Sensitivity to the set of analysed Balmer lines}
\label{set}

In order to study the sensitivity of the determined values of $T_{\rm eff}$ and $\log{g}$ to the set of analysed Balmer line profiles, we have compared the best fit results obtained when one of the Balmer lines is omitted from the analysis with the best fit results obtained from the analysis of all nine Balmer line profiles for HD~129956 (108 Vir) and HD~87901 ($\alpha$ Leo) using the Phoenix-15 grids and the Atlas9 grids. The fitting procedure have been performed with the help of FITSB2 code \citep{Napiwotzki+04} employing the aforementioned grids of stellar atmosphere models together with the respective simulated spectra.
The left panel of Fig.~\ref{fig2a} shows the differences between the values of $T_{\rm eff}$ obtained by omitting one of the Balmer lines from the analysis and the value of $T_{\rm eff}$ given in Section~\ref{grid} for HD~129956 and HD~87901 respectively. The corresponding differences for $\log{g}$ are shown in the right panel of Fig.~\ref{fig2a} for the same stars. In the case of $T_{\rm eff}$ and $\log{g}$, the zero difference corresponds to the respective values obtained using the Phoenix-15 grids and presented in Fig.~\ref{fig1} for each reference star. Meanwhile, the results shown in the Fig.~\ref{fig2a} for the Atlas9 grids are compared to the corresponding values of stellar parameters found for the reference stars in Section~\ref{grid} using the Atlas9 grids.

%Taking into account that the effective temperature of HD~129956 is close to 10000~K,
For HD~129956 (108~Vir), we can see that its best fit values of $T_{\rm eff}$ and $\log{g}$ jump up if one uses the Atlas9 grids and excludes from the analysis one of the following Balmer lines $H_{\delta}$, $H_{\epsilon}$, $H8$ or $H9$. Meanwhile, the use of the Phoenix-15 grids results in relatively small variation of $T_{\rm eff}$ and $\log{g}$ when one of these Balmer lines is excluded from the analysis.

In the case of HD~87901 ($\alpha$ Leo), %with $T_{\rm eff}$ = 11956~K
application of the Atlas9 grids or of the Phoenix-15 grids leads, under the same conditions, to relatively small variations of $T_{\rm eff}$ and $\log{g}$ that do not exceed 100~K and 0.06 respectively (see Fig.~\ref{fig2a}). The errors chosen for $T_{\rm eff}$ ($\pm$200~K) and $\log{g}$ ($\pm$0.2) therefore seem reasonable.
It appears that the use of the Phoenix-15 grids results in more stable best fit data for $T_{\rm eff}$ and $\log{g}$ with respect to the set of Balmer line profiles employed for the analysis. Meanwhile, the use of the Atlas9 grids may cause significant errors in the determination of fundamental stellar parameters if the effective temperature is close to 10000~K, depending on which set of Balmer lines is used.
%Our researches show that in the case of the use the Atlas9 grids these errors depend strongly on the set of analysed Balmer lines.

The variations of $T_{\rm eff}$ and $\log{g}$ are most sensitive to the $H_{\beta}$, $H_{\gamma}$, $H_{\delta}$, and $H_{\epsilon}$ Balmer lines when one employs the Phoenix-15 grids. It seems that this sensitivity does not vary much in the range of effective temperatures from 9700~K to 12000~K (see Fig.~\ref{fig2a}).

\subsection{The best fit results}
\label{best}

\begin{figure*}[t]
\includegraphics[width=3.25in,angle=0]{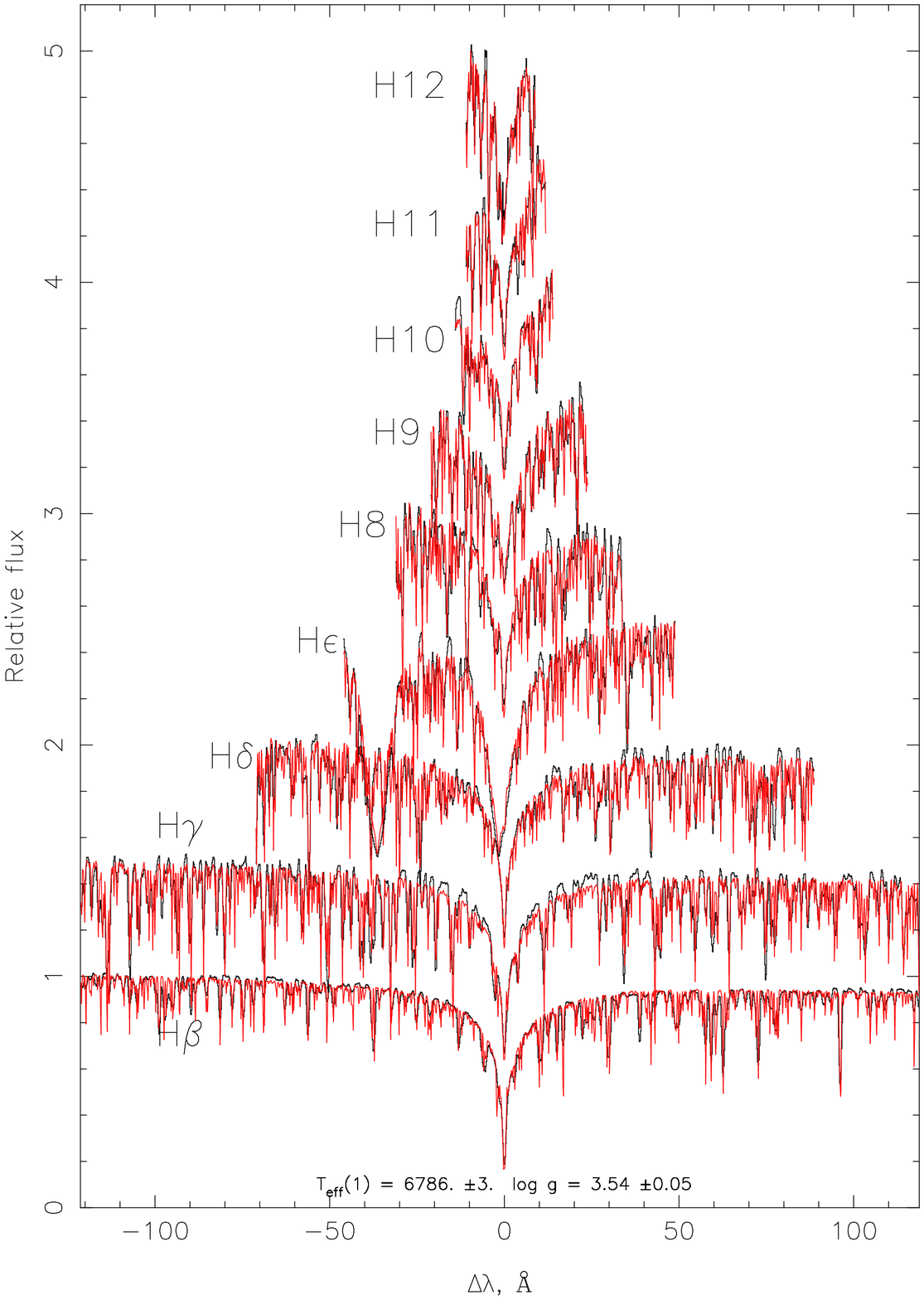}
\includegraphics[width=3.25in,angle=0]{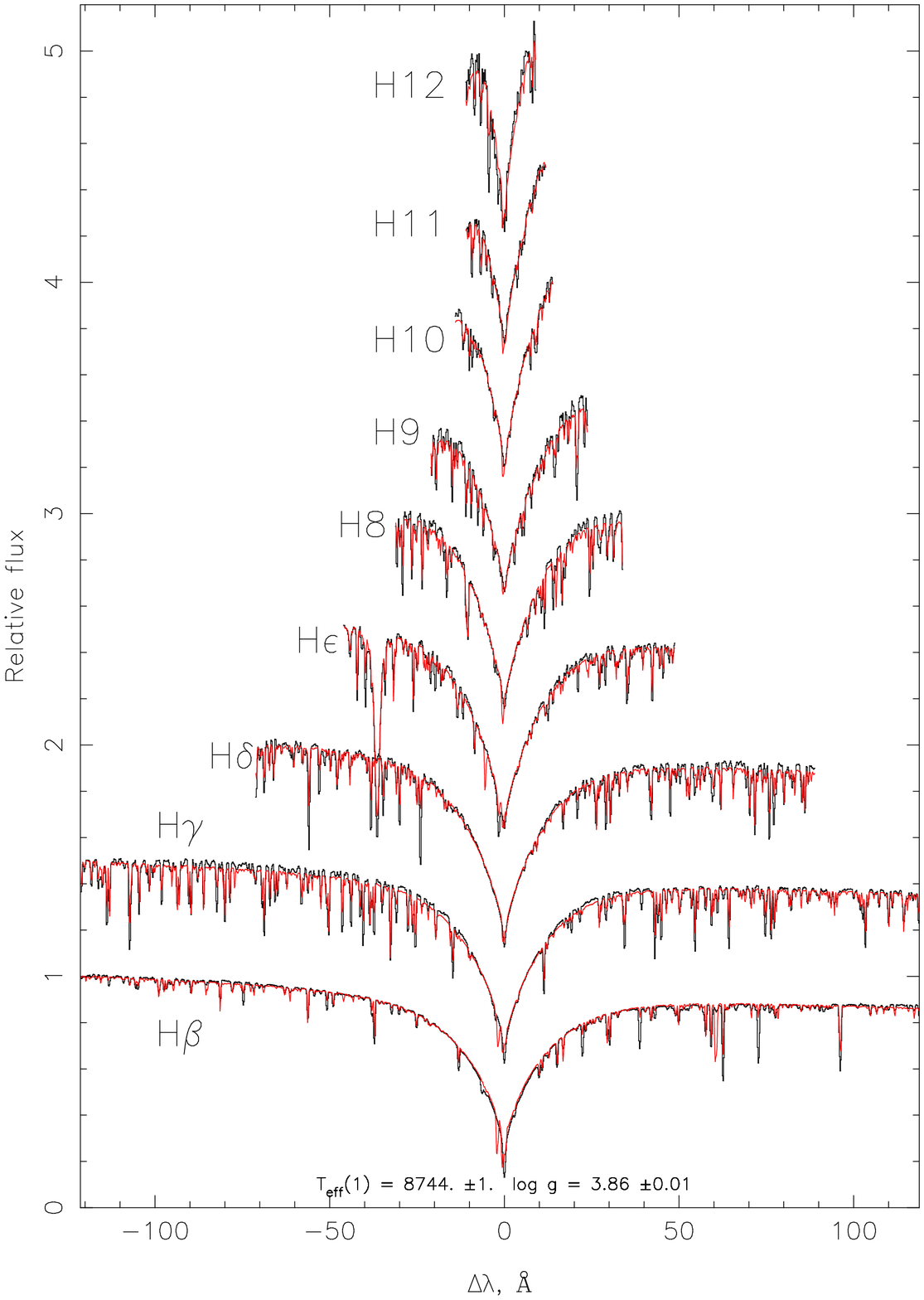}
\caption{The observed spectrum (thick line) of HD~164584 (left) and HD~23878 (right) is well fitted by synthetic spectrum (thin dotted line) that corresponds to $T_{\rm eff}$ = 6800 K, $\log{g}$ = 3.54 ($\chi^2/\nu$ = 1.16) and $T_{\rm eff}$ = 8740 K, $\log{g}$ = 3.86 ($\chi^2/\nu$ = 0.62) respectively.}
%The observed spectrum (black line) of HD95608 (left) and HD90277 (right) is well fitted by synthetic spectrum (red line) that corresponds to $T_{\rm eff}$ = 9800 K, $\log{g}$ = 3.82 %($\chi^2/\nu$ = 0.97) and  $T_{\rm eff}$ = 7250 K, $\log{g}$ = 3.65 ($\chi^2/\nu$ = 1.29) respectively.
\label{fig2}
\end{figure*}

Nine Balmer line profiles have been fitted in the observed spectra of the selected slowly rotating CP stars to find their effective temperature and surface gravity.
For each star, the fitting procedure has been performed for the metallicities [M/H]= -1.0, -0.5, 0.0, +0.5, +1.0 using the Phoenix-15 grids. Among the obtained results, the fundamental parameters associated to the fit with the smallest value $\chi^2/\nu$ are chosen.
These best fit results for effective temperature, surface gravity, radial velocity and metallicity are presented respectively in the second, third, fourth and fifth columns of Table~\ref{tab1b} together with the fit quality in the sixth column.
%The precision of the estimates for the effective temperature and the surface gravity is mentioned above.
The values of the effective temperature and the surface gravity found for these stars by other authors are given in the seventh and eighth columns respectively. Meanwhile, the previously published values for $V\sin{i}$ are presented in the ninth column.
%$V_{\rm r}$ and $V\sin{i}$ are presented in the ninth and tenth columns respectively.
Examples of best fits for \objectname{HD~23878} and \objectname{HD~164584}, and for \objectname{HD~95608} and \objectname{HD~209459} are shown in Fig.~\ref{fig2} and Fig.~\ref{fig3} respectively.

\section{Individual stars}
\label{stars}

\subsection{HD15385}

\citet{Kunzli+North98} have classified \objectname{HD~15385} (HR~723) as a A2m star having a mass $M_{\rm *}$ = 1.85$M_{\rm \sun}$ and age $\log{t}$ = 8.79$\pm$0.12. Taking into account its relatively small value of rotational velocity $V\sin{i}$ = 29 km s$^{-1}$ \citep{Royer+02}, this star was chosen for the VeSElkA project. Our estimate of the effective temperature $T_{\rm eff}$ = 8230$\pm$200~K and the surface gravity $\log{g}$ = 4.0$\pm$0.2 are in good agreement with the values found by \citet{Kunzli+North98} (see Tab.~\ref{tab1b}). %We have obtained two spectra of this star that do not show a significant line profile variability in the span of 29 days.
%Say which chemical elements are present there
Our estimate of the radial velocity for \objectname{HD~15385} (see Tab.~\ref{tab1b}) is similar to the value of $V_{\rm r}$ = 21.3 km s$^{-1}$ obtained by \citet{Wilson53}, but is significantly higher than the value $V_{\rm r}$ = 15 km s$^{-1}$ reported by \citet{Palmer+68}.

%We have obtained two Stokes IV spectra of \objectname{HD~15385} with a time span of 29 days. These spectra do not show any significant variability of the observed line profiles during this time.

%have found for this star $T_{\rm eff}$ = 8154~K and $\log{g}$ = 4.12. Our

\subsection{HD22920}

The silicon star \objectname{HD~22920} (22~Eri, HR~1121) shows a weak photometric variability with the period P = 3$^d$.95 \citep{Bartholdy88} and the presence of a rather weak longitudinal magnetic field $B_{\rm l}$ = 310$\pm$160~G \citep{Bychkov+03}. \citet{Catanzaro+99} have found $T_{\rm eff}$ = 13700~K and $\log{g}$ = 3.72 for this star. Similar parameters ($T_{\rm eff}$ = 13700~K and $\log{g}$ = 3.72) have been adopted by \citet{L+M96} to reproduce the observed spectrum of \objectname{HD~22920}. Our estimate of the effective temperature and the surface gravity (see Tab.~\ref{tab1b}) are consistent with these results. Meanwhile, the radial velocity $V_{\rm r}$ = 20.0$\pm$2.0 km s$^{-1}$ obtained in the present study seems to be higher than the previously reported values 16.4$\pm$3.1 km s$^{-1}$ (Leone \& Manfr\`{e} 1996; Wielen et al. 2000) and 17.2$\pm$0.7 km s$^{-1}$ \citep{Gontcharov2006}. %This fact suggests that \objectname{HD~22920} may be the member of a long period binary system.
From the analysis of \objectname{HD~22920} spectra, \citet{L+M96} have also found its rotational velocity $V\sin{i}$ = 30 km s$^{-1}$.

We have acquired several Stokes IV spectra for this star that show a strong variability of Si\,{\sc ii}, Ti\,{\sc ii}, Cr\,{\sc ii}, Fe\,{\sc ii} line profiles with rotational phase, while Mg\,{\sc ii} and He\,{\sc i} line profiles seem to be less variable. Our high S/N spectra (see Tab.~\ref{tab1}) reveal a weak variability of the He\,{\sc i} 5876\AA\, line profile for which \citet{Catanzaro+99} have found no variability.

%A preliminary abundance analysis has shown that Si\,{\sc ii} and Cr\,{\sc ii} increase their abundance towards the deeper atmospheric layers in the atmosphere of \objectname{HD~22920} for the all three analysed rotational phases \citep{Khalack+Poitras14}. To study the influence of the magnetic field on the vertical and horizontal stratification of chemical species in the stellar atmosphere of \objectname{HD~22920}, we need to obtain more Stokes IV spectra with high S/N ratio well spread (in phases) over the period of rotation.

\subsection{HD~23878}

Based on the measurement of the anomalously low line strength ratio Sc\,{\sc ii} 4246\AA/Sc\,{\sc ii} 4215\AA, \citet{Conti65} has suggested that \objectname{HD~23878} might be an Am star.
\objectname{HD~23878} (28~Eri, HR~1181) appears to be a low amplitude variable star with a period of about 7$^d$.17 \citep{Mathys+85}
%In the catalog of \citet{R+M09} this object is classified as a star of spectral class A1
having a rotational velocity $V\sin{i}$ = 24 km s$^{-1}$ \citep{Royer+02}.
Meanwhile, \citet{Abt+Morrell95} have reported $V\sin{i}$ = 18 km s$^{-1}$ for this star.
\objectname{HD~23878} has not previously been studied spectroscopically in detail and the fitting of the Balmer line profiles (see right panel at the Fig.~\ref{fig2}) results in $T_{\rm eff}$ = 8740$\pm$200~K and $\log{g}$ = 3.86$\pm$0.20. Our estimate of radial velocity %$V_{\rm r}$ = 29.5$\pm$1.0 km s$^{-1}$
is in agreement with the value $V_{\rm r}$ = 28.4$\pm$0.5 km s$^{-1}$ obtained by \citet{Gontcharov2006} (see Tab.~\ref{tab1b}).

%We have obtained two Stokes IV spectra of \objectname{HD~23878} separated by 53 days. The numerous observed line profiles do not show any significant variability during this period of time.

%\subsection{HD~24712}
%Ap star \objectname{HD~24712} shows a weak photometric variability with the rotational period P = 12$^d$.46 \citep{ESA97}. The observed longitudinal magnetic field also varies with the same period P = 12$d$.45812(19) reaching it maximum $B_{\rm l}$ = 1.15$\pm$0.10kG (Ryabchikova et al. 2005, 2007; Rusomarov et al. 2013).

%May be do not report this star: log g is different than the value obtained by Ryabchikova et al. (2005)

\subsection{HD~53929}

\objectname{HD~53929} is a HgMn star with mild but significant enhancement of manganese abundance $\log{N_{\rm Mn}/N_{\rm H}}$ = -5.85$\pm$0.20 \citep{Smith+Dworetsky93} and with a mercury abundance $\log{N_{\rm Hg}/N_{\rm H}}$ = -9.90$\pm$0.20 at the level of normal stars \citep{Smith97}. This star has also much lower abundance of chromium, cobalt and nickel as compared to their solar abundances \citep{Smith+Dworetsky93}. The analysis of Balmer line profiles in the single spectrum available for \objectname{HD~53929} results in an effective temperature %$T_{\rm eff}$ = 13950$\pm$200~K
which is in good agreement with the temperature obtained by \citet{Smith+Dworetsky93} (see Tab.~\ref{tab1b}). Meanwhile, our estimate of the surface gravity $\log{g}$ = 3.90$\pm$0.20 seems to be a little higher than the value $\log{g}$ = 3.60$\pm$0.25 reported by \citet{Smith+Dworetsky93}.

\citet{Royer+02} have obtained $V\sin{i}$ = 25 km s$^{-1}$ for \objectname{HD~53929}. Its radial velocity increased over the last 50 years going from 6.1 km s$^{-1}$ (Evans 1967; Hube 1970), to 11.2$\pm$1.3 km s$^{-1}$ \citep{Gontcharov2006} and to 15.0$\pm$1.0 km s$^{-1}$ found in this study. Is seems that \objectname{HD~53929} may be a member of a long periodic double system.

\subsection{HD~68351}

In the catalog of \citet{R+M09}, this object is classified as a star of spectral class A0 with the strong lines of Si and Cr. According to \citet{Abt+Morrell95}, the rotational velocity of \objectname{HD~68351} is $V\sin{i}$ = 25 km s$^{-1}$, while \citet{Royer+02} and \citet{Auriere+07} have reported $V\sin{i}$ = 33 km s$^{-1}$. Its rotational period $P= 4^{d}.116$ was determined by \citet{Stepien68}.
In this study, we present the results for its fundamental parameters $T_{\rm eff}$ = 10080$\pm$200~K and $\log{g}$ = 3.22$\pm$0.20 based on the analysis of Balmer line profiles which are in agreement with the data previously published by \citet{Auriere+07} for this star. Our estimate of the radial velocity seems to be a little smaller than the value $V_{\rm r}$ = 19.9 km s$^{-1}$ obtained by \citet{Evans67} (see Tab.~\ref{tab1b}).

Taking into account the high values of luminosity $L_{\rm *}=466\pm225 L_{\sun}$ and radius $R_{\rm *}=6.6\pm2.4 R_{\sun}$ of \objectname{HD~68351} \citep{Auriere+07} and its low surface gravity $\log{g}$ = 3.22$\pm$0.20, this star is beyond the end of the main sequence and most probably belongs to the III luminosity class. \citet{Auriere+07} have detected the presence of a longitudinal magnetic field $B_{l} = 325 \pm 45$ G from the analysis of HD~68351 LSD profiles.

For this star, we have obtained two Stokes IV spectra with a time gap of 2 days. The Stokes I spectrum of \objectname{HD~68351} appears to be strongly variable, while the Stokes V does not show signatures of the presence of a strong magnetic field. This is consistent with the results of \citet{Auriere+07} that have observed HD~68351 15 times, but have found the magnetic field signatures in only 5 spectra.
A preliminary analysis of the two obtained spectra results in slightly different values of radial velocity $V_{\rm r}$ = 18.1$\pm$1.0 km s$^{-1}$ and 14.7$\pm$1.0 km s$^{-1}$ that can be an indicator of possible binarity of \objectname{HD~68351}.

\begin{figure*}[t]
\includegraphics[width=3.25in,angle=0]{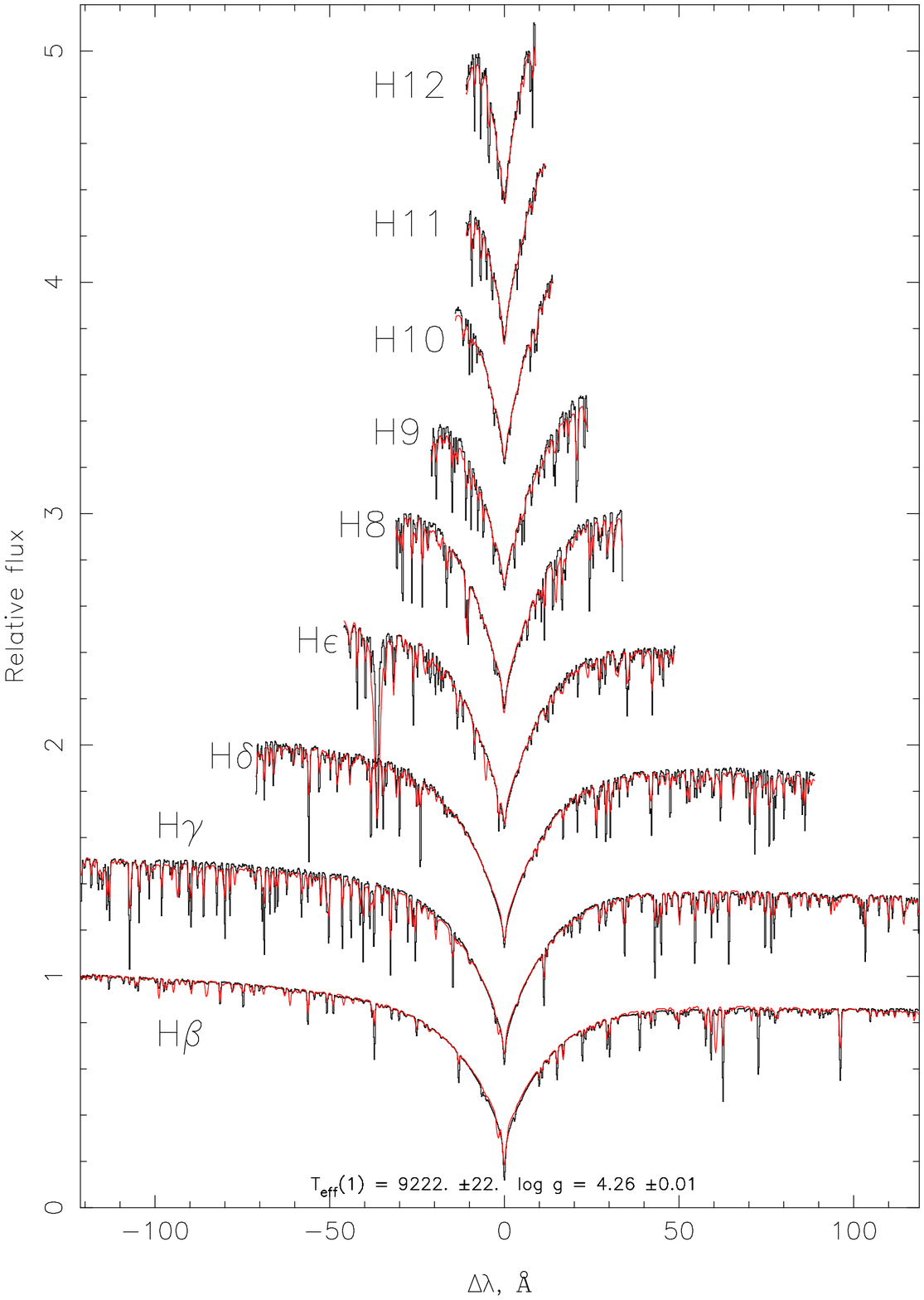}
\includegraphics[width=3.25in,angle=0]{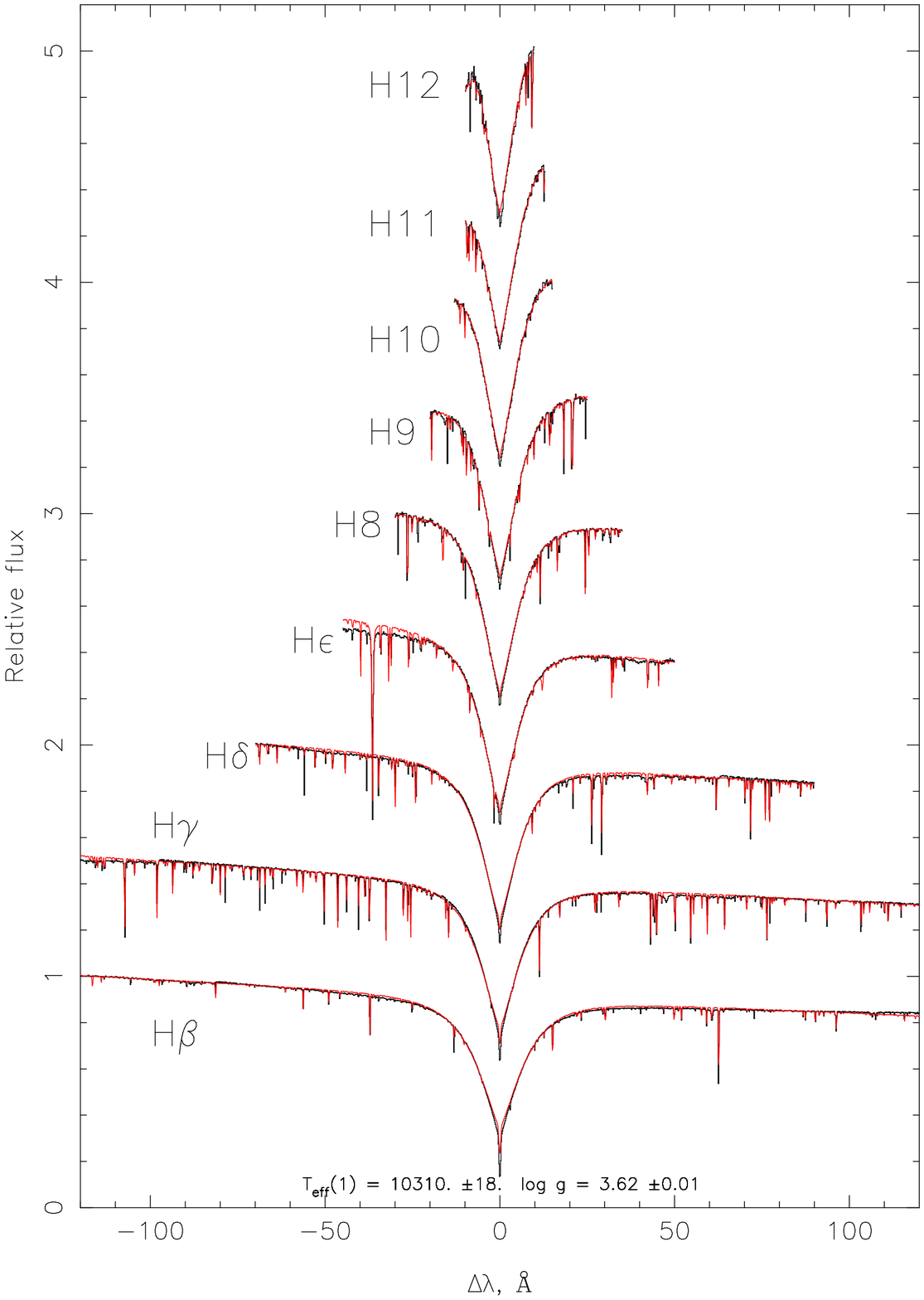}
\caption{The observed spectrum (thick line) of HD~95608 (left) and HD~209459 (right) is well fitted by synthetic spectrum (thin dotted line) that corresponds to $T_{\rm eff}$ = 9200 K, $\log{g}$ = 4.26 ($\chi^2/\nu$ = 0.59) and $T_{\rm eff}$ = 10310 K, $\log{g}$ = 3.62 ($\chi^2/\nu$ = 0.97) respectively. }
\label{fig3}
\end{figure*}

\subsection{HD~71030}

\citet{Cenarro+07} have reported \objectname{HD~71030} (25~Cnc, HR~3299) to be a main-sequence star of spectral class F6.
Based on the analysis of a low resolution spectrum, \citet{Balachandran90} has reported higher than solar lithium abundance $\log{N_{\rm Li}/N_{\rm H}}$ = -9.37 and slightly underabundant iron $\log{N_{\rm Fe}/N_{\rm H}}$ = -4.80$\pm$0.05 for this star. Meanwhile, \citet{Khalack+13} have found nearly solar abundance of Fe\,{\sc ii} $\log{N_{\rm Fe}/N_{\rm H}}$ = -4.43$\pm$0.34, Cr\,{\sc ii} $\log{N_{\rm Cr}/N_{\rm H}}$ = -6.27$\pm$0.09 and Ni\,{\sc i} $\log{N_{\rm Ni}/N_{\rm H}}$ = -5.77$\pm$0.26.
They have also found that Fe, Cr and Ni are uniformly distributed with respect to optical depth in the atmosphere of this star.
%\objectname{HD~71030} is included into the catalogue of \citet{R+M09} as a chemically peculiar star.

Our estimation of the surface gravity for \objectname{HD~71030} is the same as the value $\log{g}$ = 4.03$\pm$0.05 published by \citet{Prugniel+11} (see Tab.~\ref{tab1b}), while our estimation of the effective temperature $T_{\rm eff}$ = 6780$\pm$200~K seems to be slightly higher as compared to the results of \citet{Prugniel+11} (but within the given error bars). The value for the radial velocity determined in this work %$V_{\rm r}$ = 38.1$\pm$1.0 km s$^{-1}$
is in agreement with the previously obtained results of 37.4 km s$^{-1}$ \citep{Nordstreom+04} and 37.1$\pm$0.4 km s$^{-1}$ \citep{Gontcharov2006}. The estimate of $V\sin{i}$  = 9$\pm$2 km s$^{-1}$ \citep{Khalack+13} is also in good agreement with the value $V\sin{i}$  = 8 km s$^{-1}$ provided by Balachandran (1990).

%\objectname{HD~71030} has been observed twice with a time gap of 2 days. There is no significant variability of the observed line profiles during this period. Preliminary abundance analysis has shown that Fe, Cr and Ni are uniformly distributed with respect to optical depth in the atmosphere of this star \citep{Khalack+13}.

\subsection{HD~83373}

\objectname{HD~83373} (34~Hya, HR~3832) is reported by \citet{R+M09} as a chemically peculiar star of spectral class A1 that has an enhanced silicon abundance.
Royer et al. (2002, 2007) have measured its rotational velocity $V\sin{i}$  = 28 km s$^{-1}$. This star has not been previously studied spectroscopically and we provide here first estimates of its effective temperature $T_{\rm eff}$ = 9800$\pm$200~K and surface gravity $\log{g}$ = 3.81$\pm$0.20. Our estimate of the radial velocity for \objectname{HD~83373} (see Tab.~\ref{tab1b}) is in good agreement with the value $V_{\rm r}$ = 26.9$\pm$0.5 km s$^{-1}$ reported by \citet{Gontcharov2006}.

%We have observed \objectname{HD~83373} twice with a time gap of 65 days. The Stokes I spectra show numerous spectral lines some of which appear to be slightly variable. We have not detected any signatures of a strong magnetic field in its Stokes V spectra.

\subsection{HD~90277}

Abundance analysis of \objectname{HD~90277} (HR~4090) has been performed by \citet{Berthet90} who has found a strong overabundance of Y, Zr and Ba, while Ti, Cr, Mn, Fe and Ni appear to be slightly overabundant. Royer et al. (2002, 2007) have reported $V\sin{i}$  = 34 km s$^{-1}$ for \objectname{HD~90277}. Our estimates of
%$T_{\rm eff}$ = 7250$\pm$200~K  and $\log{g}$ = 3.62$\pm$0.20
the effective temperature and the surface gravity are similar to the values obtained by \citet{Berthet90} taking into account the error bars (see Tab.~\ref{tab1b}).
The value for the radial velocity determined in this work %$V_{\rm r}$ = 14.5%$\pm$1.0 km s$^{-1}$
agrees well with the previously obtained result $V_{\rm r}$ = 13.7 km s$^{-1}$ \citep{Evans67} and 13.7$\pm$0.6 km s$^{-1}$ \citet{Gontcharov2006}.

%This star has been observed twice with a time gap of 65 days. The observed Stokes I spectra show no variability of line profiles during this time. {\bf The} Stokes V spectra show no signatures of a strong magnetic field.

\subsection{HD~95608}

\citet{Cowley+69} have classified \objectname{HD~95608} (60~Leo, HR~4300) as a CP star of spectral class A1m.
%, while \citet{R+M09} have reported for this stars spectral class A0-A4.
From the fitting of Balmer line profiles of \objectname{HD~95608} (see left panel of Fig.~\ref{fig3}), we have found its effective temperature $T_{\rm eff}$ = 9200$\pm$200~K and surface gravity $\log{g}$ = 4.25$\pm$0.20 (see Tab.~\ref{tab1b}).
The value for radial velocity determined in this work $V_{\rm r}$ = -10.4$\pm$1.0 km s$^{-1}$ agrees well with the previously obtained results -10.1 km s$^{-1}$ \citep{Wielen+00} and -11.1$\pm$0.7 km s$^{-1}$ \citep{Gontcharov2006}. The rotational velocity $V\sin{i}$  = 17.2$\pm$2.0 km s$^{-1}$ obtained by \citet{Khalack+13} for \objectname{HD~95608} seems to be smaller than the value $V\sin{i}$  = 21 km s$^{-1}$ reported by \citet{Royer+02}, but higher than $V\sin{i}$ = 13 km s$^{-1}$ published by \citet{Abt+Morrell95}.

Preliminary abundance analysis has shown that Ti\,{\sc ii} is slightly underabundant, while Fe\,{\sc i}, Fe\,{\sc ii} are overabundant in this star as compared to their solar abundance \citep{Khalack+13}. This same study has also found that iron abundance appears to be vertically stratified in the stellar atmosphere of \objectname{HD~95608}.

\subsection{HD~97633}

\citet{R+M09} have classified \objectname{HD~97633} ($\theta$~Leo, HR~4359) as a CP star of spectral class A2 that has enhanced Sr and Eu abundances \citep{Hill95}. \citet{Royer+02} have found $V\sin{i}$  = 23 km s$^{-1}$ for this star.
%\citet{Bychkov+03} have reported for \objectname{HD~97633} presence of a significant longitudinal magnetic field with strength $B_{\rm l}$=53.6$\pm$34.1 G.
Our estimates of its effective temperature and surface gravity (see Tab.~\ref{tab1b}) are in good agreement with the results obtained by \citet{Koleva+Vazdekis12}. The radial velocity $V_{\rm r}$ = 8.2$\pm$1.0 km s$^{-1}$ found in this work is consistent with the value $V_{\rm r}$ = 7.3 km s$^{-1}$ reported by \citet{Gontcharov2006}.

%We have accumulated 9 Stokes IV spectra of this star obtained over a period of 66 days. The Stokes I spectra show significant variation of line profiles of many chemical species during this period, while the Stokes V spectra show no signatures of a strong magnetic field.

\subsection{HD~110380}

\objectname{HD~110380} (HR~4826) is in a binary system with \objectname{HD~110379} with an orbital period P=171.4yr and a separation $a$=3.75$^{\prime\prime}$. \objectname{HD~110380} has $V\sin{i}$  = 23 km s$^{-1}$ \citep{Royer+02}
%and rotational period $P_{\rm rot}$ = 0.$^d$29 \citep{B+T86}.
%(Rotational period is deduced from the enhanced lithium abundance \citep{B+T86})
%\objectname{HD~110380 is a lithium reach CP star \citep{B+T86}.
and an enhanced lithium abundance \citep{B+T86}.
Our estimate of its effective temperature seems to be a little higher than the value $T_{\rm eff}$ = 6720~K obtained by \citet{B+T86}, but the surface gravity is in good agreement with the value $\log{g}$ = 4.2 they found (see Tab.~\ref{tab1b}).
The value for the radial velocity determined in this work $V_{\rm r}$ = -17.6$\pm$1.0 km s$^{-1}$ is consistent with the value $V_{\rm r}$ = -19.5 km s$^{-1}$ obtained by \citet{Nordstreom+04}.

%We have obtained two Stokes IV spectra of \objectname{HD~110380} with a gap of 10 days. The spectra are not contaminated by lines of the secondary component and the line profiles observed in Stokes I spectra only show small variability of their intensity over the period of observation.
%Delete this star? rapid rotation? Observed spectrum are not contaminated by the primary component!!

\subsection{HD~116235}

\citet{Cowley+69} have classified \objectname{HD~116235} (HR~5040, 64~Vir) as a CP star of spectral class A2m. %, while \citet{R+M09} have reported for this stars spectral class A2-A8.
The effective temperature $T_{\rm eff}$ = 8900$\pm$200~K obtained in this study for \objectname{HD~116235} %(see Tab.~\ref{tab1b})
appears to be higher than the previously published values 8570~K \citep{Erspamer+North03} and 8373~K \citep{Ammler+Reiners12}.
Meanwhile, the surface gravity obtained here is similar to $\log{g}$ = 4.23 reported by \citet{Erspamer+North03}. The estimate of $V\sin{i}$  = 20$\pm$2.0 km s$^{-1}$ \citep{Khalack+13} is in a good agreement with the value 19.3$\pm$1.0 km s$^{-1}$ obtained by \citet{Ammler+Reiners12} for this star. The radial velocity $V_{\rm r}$ = -10.3$\pm$1.0 km s$^{-1}$ obtained in this study is consisted with the previously published values  -10.2 km s$^{-1}$  \citep{Wilson53} and -8.4$\pm$2.3 km s$^{-1}$ \citep{Gontcharov2006}.

A preliminary abundance analysis of this star shows an enhanced abundance of  Ni\,{\sc i}, Fe\,{\sc i}, Fe\,{\sc ii}, Cr\,{\sc i} and Cr\,{\sc ii} \citep{Khalack+13} in agreement with the results of \citet{Erspamer+North03}, who have also reported a strong overabundance of Sr, Y and Ba. Moreover, \citet{Khalack+13} have also found signatures of vertical stratification of iron abundance in the atmosphere of \objectname{HD~116235}.

%For \objectname{HD~116235}, we have obtained two Stokes IV spectra with a time gap of two days. The line profiles observed in Stokes I spectra do not show significant variability during this period. The preliminary abundance analysis of this star shows an enhanced abundance of  Ni\,{\sc i}, Fe\,{\sc i}, Fe\,{\sc ii}, Cr\,{\sc i} and Cr\,{\sc ii} \citep{Khalack+13} in agreement with the results of \citet{Erspamer+North03}, who have also reported a strong overabundance of Sr, Y and Ba. Moreover, \citet{Khalack+13} have found signatures of vertical stratification of iron abundance in the atmosphere of \objectname{HD~116235}.

%\begin{figure*}
%\includegraphics[width=2.4in,angle=-90]{HD164584_1656316_Mg2_4481I.eps}
%\includegraphics[width=2.4in,angle=-90]{HD164584_1656316_Mg2_4481I.eps}
%\includegraphics[width=2.4in,angle=-90]{HD164584_1650322_Mg2_4481I.eps}
%\includegraphics[width=2.4in,angle=-90]{HD164584_1650322_Mg2_4481I.eps}
%\caption{The fitted Mg\,{\sc ii} 4481\AA\, line profile observed in the spectra of \objectname{HD~164584} with $\chi^2/\nu$ = 2.44 and 0.07 respectively.
%}
%\label{fig4}
%\end{figure*}

\subsection{HD~164584}

\citet{R+M09} have reported that \objectname{HD~164584} (7~Sgr, HR~6724) has a spectral class A6-F4.
%, while Uesugi \& Fucuda (1970) have published its $V\sin{i}$  = 24 km s$^{-1}$ ---> it is not published.
%Vsini?? from FeII line 4923
From the analysis of Balmer line profiles (see left panel of Fig.~\ref{fig2}), we have obtained $T_{\rm eff}$ = 6800$\pm$200~K and $\log{g}$ = 3.54$\pm$0.20.

%{\bf Preliminary analysis of the Mg\,{\sc ii} 4481\AA\, line profile, has resulted in its rotational velocity $V\sin{i}$  = 30.2$\pm$1.5 km s$^{-1}$ that is close to the value 24 km s$^{-1}$ derived by \citet{Uesugi+Fukuda70} for \objectname{HD~164584}.}
Our estimate of radial velocity (see Tab.~\ref{tab1b}) is in good agreement with the value $V_{\rm r}$ = -11.2$\pm$1.6 km s$^{-1}$ obtained by \citet{Gontcharov2006} for this star.

%We have obtained two Stokes IV spectra of \objectname{HD~164584} with a time gap of 76 days. The line profiles in Stokes I spectra show some weak variability of intensity during this period.

%\subsection{HD~170973}
%The value of radial velocity determined in this work $V_{\rm r}$ = -10.5%$\pm$1.0
%km s$^{-1}$ differs significantly from the previously obtained data -8.3 km s$^{-1}$ \citep{Evans67} and  -13.8$\pm$0.8 km s$^{-1}$ \citep{Gontcharov2006}.
%This may be an indicator that the star is a member of double system.

\subsection{HD~186568}

\objectname{HD~186568} (HR~7512) %is member of a double system and
belongs to the normal B-type stars (B9 II) \citep{Hubrig+Castelli01}. Nevertheless, it was included by \citet{Renson+Manfroid09} to the list of chemically peculiar stars. From the analysis of equivalent widths of iron line profiles \citet{Hubrig+Castelli01} have found its solar abundance in the atmosphere of this star.

The effective temperature obtained for \objectname{HD~186568} in this study is smaller than the value derived by \citet{Hubrig+Castelli01}, while our value for the surface gravity is almost the same as the one determined by these authors %\citet{Hubrig+Castelli01}
(see Tab.~\ref{tab1b}). Our estimate for radial velocity is close to $V_{\rm r}$ = -8.8$\pm$0.1 km s$^{-1}$ reported by \citet{Gontcharov2006} and to the value -8.0$\pm$1.4 km s$^{-1}$ reported by \citet{Morrell+Abt92} who have used this star as a radial velocity standard.

%{\bf From the analysis of the Mg\,{\sc ii} 4481\AA\, line profile, we have found its rotational velocity $V\sin{i}$  = 17.5$\pm$1.5 km s$^{-1}$ that is very similar to the one found by \citet{Hubrig+Castelli01} and by \citet{Fekel03}. }
%A preliminary abundance analysis has shown that the magnesium in this star appears to be slightly overabundant [Mg/H]=+0.69$\pm$0.07 with respect to its solar abundance.

%\citet{Hubrig+Castelli01} and \citet{Fekel03} have reported that \objectname{HD~186568} is normal B-type star and we plan to use it as a reference star for our project. Our estimates of $T_{\rm eff}$ = 11070$\pm$200~K and $\log{g}$ = 3.44$\pm$0.20 for this star are in a good accordance with the results obtained by \citet{Hubrig+Castelli01} (see Tab.~\ref{tab1b}). For \objectname{HD~186568}, we have obtained two Stokes IV spectra separated in time by 6 days. The Stokes I spectra show no significant variability during this period, and the Stokes V spectra show no signatures of a strong magnetic field. }

\subsection{HD~209459}

\objectname{HD~209459} (21~Peg, HR~8404) is a normal B-type star (B9.5 V) which is often used as a comparison star \citep{Dworetsky+Budaj00} because of its sharp-lined spectrum with $V\sin{i}$  = 4 km s$^{-1}$ \citep{Smith92} and absence of chemical abundance peculiarities. %and according to \citet{Cowley+Aikman80} it does not show chemical peculiarities and magnetic field signatures.
Based on the analysis of equivalent widths of iron line profiles in the spectrum of \objectname{HD~209459}, \citet{Hubrig+Castelli01} have found an iron abundance that is close to its solar value. We also aim to use this object as a comparison star in the VeSElkA project.

Our estimate of effective temperature and surface gravity (see Tab.~\ref{tab1b}) obtained from the analysis of Balmer line profiles (see right panel of Fig.~\ref{fig3}) is in agreement with the previously published results $T_{\rm eff}$ = 10455$\pm$400~K, $\log{g}$ = 3.52$\pm$0.15 of \citet{Hubrig+Castelli01}, but smaller than the results  $T_{\rm eff}$ = 11015$\pm$301~K, $\log{g}$ = 3.99$\pm$0.12 reported by \citet{Prugniel+11}. The radial velocity found in this study (see Tab.~\ref{tab1b}) is also in a good agreement with the value $V_{\rm r}$ = 0 km s$^{-1}$ reported by \citet{Smith92}.

%For \objectname{HD~209459}, we have obtained two Stokes IV spectra separated in time by 6 days. The observed line profiles seem to be non-variable during the period of observation in Stokes I spectra, while the Stokes V spectra show no signatures of a strong magnetic field.

\subsection{HD~223640}

\citet{R+M09} have reported \objectname{HD~223640} (108~Aqr, HR~9031) as a CP star of spectral class B9. It shows an overabundance of Si, Sr and Cr \citep{North+92}. \citet{Bailey+Landstreet13} have found $V\sin{i}$  = 31$\pm$3 km s$^{-1}$ for this star and that Cr and Si abundances exceed their respective solar values more than by 1 dex. Analysis of polarization in Balmer lines reveals the presence of a longitudinal magnetic field $B_{\rm l}$ = 643$\pm$218~G \citep{Bychkov+03} which varies with period $P$ = 3$^d.$73 \citep{North+92}. The same period $P$ = 3$^d$.735239$\pm$0$^d$.000024 has been found by \citet{North+92} from photometric variability of \objectname{HD~223640}.

The effective temperature and the surface gravity obtained in this study (see Tab.~\ref{tab1b}) are in good agreement with the results of \citet{Prugniel+11}, but seem to be slightly different (but still inside the error bars) with respect to the values  $T_{\rm eff}$ = 12300$\pm$500~K, $\log{g}$ = 4.4$\pm$0.2 reported by \citet{Bailey+Landstreet13}. Our estimate of the radial velocity seems to be larger (but still within the error bars) than the value $V_{\rm r}$ = 12.7$\pm$2.8 km s$^{-1}$ reported by \citet{Gontcharov2006}.

%We have obtained two Stokes IV spectra of \objectname{HD~223640} within a time gap of 10 days. The observed line profiles are strongly variable in the both Stokes I and V spectra due to the presence of magnetic field of dipolar structure \citep{North+92} that cause the magnetic widening of the line profiles. Taking into account that angle between the axis of magnetic dipole and the rotational axis is $\beta=36^{\circ}$ \citep{North+92}, for different rotational phases an observer sees different configuration of visible magnetic field and the magnetic widening of the line profiles is therefore variable in time.

\section{Discussion}
\label{summary}
Based on the catalogue of Ap, HgMn and Am stars \citep{Renson+Manfroid09} and the available measurement of rotational velocities of CP stars (Royer et al. 2002, 2007) we have compiled a list of CP stars (see Table~\ref{tab1}) suitable for search of signatures of abundance stratification of chemical species with respect to optical depth in their atmospheres.
These stars have been recently observed in the frame of VeSElkA project with the ESPaDOnS at CFHT. The method developed for the analysis of vertical stratification of chemical abundance \citep{Khalack+Wade06} allows to determine the slope of abundance change relative to optical depth.

In order to study the vertical stratification of chemical abundance one needs to adopt an appropriate model for the  stellar atmosphere and to determine its parameters like effective temperature, surface gravity and metallicity. We were able to derive these parameters for the listed stars (see Table~\ref{tab1b}) based on the analysis of nine Balmer lines (see, for example, Fig.~\ref{fig2} and Fig.~\ref{fig3}) using the FITSB2 code \citep{Napiwotzki+04} that employs a grid of theoretical fluxes calculated for different values of $T_{\rm eff}$, $\log{g}$ and metallicity. Taking into account that the resolution R=65000 of the ESPaDOnS spectra, we have calculated a new library of synthetic spectra with a similar spectral resolution to properly fit the profiles of nine Balmer lines of stars observed in the frame of the project VeSElkA (see Section~\ref{fit}). Grids of stellar atmosphere models and corresponding fluxes have been calculated with the PHOENIX code \citep{Hauschildt+97} for 5000~K $\leqslant T_{\rm eff} \leqslant$ 15000~K and 3.0 $\leqslant \log{g} \leqslant$ 4.5 for the metallicities [M/H]= -1.0, -0.5, +0.5, +1.0, +1.5.

%!! Change it

We have used the spectra of 108~Vir (HD~129956) and $\alpha$ Leo (HD~87901) obtained with ESPaDOnS in the spectropolarimetric mode to verify the accuracy of our grids of synthetic fluxes. The results obtained for for these reference stars are close to the previously published data for their effective temperature and surface gravity (see Section~\ref{grid}).

%{\bf The fundamental parameters of stellar atmosphere of these reference stars obtained from the best fit of the nine Balmer line profiles using the Phoenix-15 grids have been compared with the best fit results of the same set of Balmer lines using the Atlas9 and Phoenix-16 grids. }
%They are also in good agreement with the results obtained for Vega using the grid of theoretical fluxes calculated for the metallicity [M/H]=-0.5 by \citet{Husser+13} using version 16 of the PHOENIX code \citep{Hauschildt+97}.

In order to study the sensitivity of the determined values of $T_{\rm eff}$ and $\log{g}$ to the set of Balmer lines used, we have compared the best fit results obtained when one of the Balmer lines is omitted from the analysis with the best fit results obtained from the analysis of all nine Balmer line profiles for the reference stars using the Phoenix-15 and Atlas9 grids. We have found that the use of the Atlas9 grids may produce some ambiguity in the determination of fundamental stellar parameters if the effective temperature is close to 10000~K depending on which set of Balmer lines is used. Employing our Phoenix-15 grids for simulations with different sets of Balmer lines, we have shown that the estimates of $T_{\rm eff}$ and $\log{g}$ are most sensitive to the $H_{\beta}$, $H_{\gamma}$, $H_{\delta}$, and $H_{\epsilon}$ Balmer lines. This sensitivity does not change significantly in the range of effective temperatures from 9700~K to 12000~K (see Fig.~\ref{fig2a}).

%From a comparison of the best fit results obtained for the reference stars during the fitting procedure of nine Balmer lines using the Phoenix-15, Phoenix-16 and Atlas9 grids and from the variation of $T_{\rm eff}$ and $\log{g}$ when one of the Balmer lines is omitted from the analysis employing the Phoenix-15 and Atlas9 grids,
We roughly estimate the uncertainties $\pm$200 K and $\pm$0.2 respectively for the values of effective temperature and surface gravity. The relatively small variabilities found for $T_{\rm eff}$ and $\log{g}$ when using subsets of Balmer lines (see Subsection~\ref{set}), seem to show that this choice for the errors is reasonable.
It should be also noted that \citet{Husser+13} have calculated grids of theoretical fluxes for $T_{\rm eff}$ below 12000~K. Therefore we can not use those grids to determine $T_{\rm eff}$ and $\log{g}$ of all the stars selected for this study because some of them are much hotter than 12000~K (see Table~\ref{tab1b}).
Meanwhile, our grids of synthetic spectra provide similar results as the grids of \citet{Husser+13} for the selected stars with $T_{\rm eff}$ below 12000~K.
%and can be applied to the stars with $T_{\rm eff}$ up to 15000~K.
Since our grids go up to $T_{\rm eff}$ = 15000~K, they can be applied for all of the stars in our sample.

Our final results for effective temperature and surface gravity obtained for twelve of the stars presented in Table~\ref{tab1b} are consistent with the previously published data. For four other stars (HD~23878, HD~83373, HD~95608 and HD~164584), this study gives the first estimates of their $T_{\rm eff}$ and $\log{g}$.
%The parameters of stellar atmosphere model and the preliminary results of abundance analysis of HD~95608 have been discussed by \citet{Khalack+13}.

It should be noted that some stars observed in the frame of the project VeSElkA have low $T_{\rm eff}$ (see Table~\ref{tab1b}) and won't have vertical stratification of chemical species due to the presence of mixing due to convection (Richer et al. 2000; Richard et al. 2001). Nevertheless, they provide the possibility to verify the validity of the method applied to analyse the vertical stratification of chemical abundances \citep{Khalack+Wade06} and to determine their abundance peculiarities. For instance, \citet{Khalack+13} found no signatures of vertical stratification in \objectname{HD~71030} and that the abundance of the analysed chemical species is close to their solar abundance. Our sample of analysed stars  includes the reference stars \objectname{HD~186568} and \objectname{HD~209459} known to be a normal B-type stars \citep{Dworetsky+Budaj00} without abundance peculiarities \citep{Hubrig+Castelli01}. These stars can also be used to verify the applied method and also to test the results for average abundances as well. The abundance analysis for all stars, even those where no stratification is found or expected, will be presented in upcoming papers. This analysis may be useful to confirm or disprove their CP type classification.

\citet{Khalack+13} have already analysed several of the selected stars with the aim to detect vertical stratification of chemical abundances in their atmospheres and found signatures of vertical stratification of iron in \objectname{HD~95608} and \objectname{HD~116235}.
%{\bf The vertical stratification of silicon and chromium is found in the atmosphere of \objectname{HD~22920} by \citet{Khalack+Poitras14}.}
A detailed abundance analysis for the other stars of our sample is underway. We also plan to add more suitable stars to the project VeSElkA.
%with precision sufficient for verification of the stratification profiles predicted by the stratified version of the PHOENIX code \citep{LeBlanc+10}. The
A large amount of observational data relative to the vertical abundance stratification of chemical species obtained for CP stars with different effective temperatures will allows us to search for a dependence of the vertical abundance stratification relative to the effective temperature, similar to the one that we have found for BHB stars (LeBlanc et al. 2010; Khalack et al. 2010). Our results will also be useful for comparison with theoretical modelling of vertical abundance stratification in stellar atmospheres (i.e. LeBlanc et al. 2009; Stift \& Alecian 2012) and to better understand the diffusion process in the atmospheres of certain CP stars.

\section*{Acknowledgments}

We are sincerely grateful to Ralf Napiwotzki and to John Landstreet for fruitful discussions and suggestions that lead to a significant improvement of this work.
We thank Facult\'{e} des \'{E}tudes Sup\'{e}rieures et de la Recherche de l'Universit\'{e} de Moncton and NSERC for research grants.
%Prof. E. Landi Degl'Innocenti and Prof. F. LeBlanc for fruitful discussions and significant help in this research. J.D.L. acknowledges financial support from the Natural Sciences and Engineering Research Council of Canada. We are also grateful to the referee, who has carefully read the paper, and has made a number of very useful comments and suggestions which have improved the paper significantly.
The calculations have been carried out on the supercomputer {\it briarree} of l'Universit\'{e} de Montr\'{e}al, under the guidance of Calcul Qu\'{e}bec and Calcul Canada. The use of this supercomputer is funded by the Canadian Foundation for Innovation (CFI), NanoQu\'{e}bec, RMGA and Research Fund of Qu\'{e}bec - Nature and Technology (FRQ-NT).
%\vspace{0.2cm}
%\noindent
This paper has been typeset from a \TeX/\LaTeX\, file prepared by the authors.

\label{lastpage}

\end{document}